\documentclass[aps,prl,twocolumn, superscriptaddress]{revtex4-2}
\usepackage[pdftex]{graphicx} 
\usepackage{placeins,float,textcomp,fontenc,extramarks,fancyhdr, enumitem, import, ctable, mathtools}
\usepackage{xcolor,soul}
\usepackage{physics,braket}
\usepackage{fullpage,makecell}

\begin{document}

\title{Magnetic tuning of tunnel coupling between InAsP double quantum dots in InP nanowires}

\author{Jason Phoenix} 
\affiliation{National Research Council Canada, Ottawa, Ontario, K1A0R6, Canada }
\affiliation{Department of Physics and Astronomy, University of Waterloo, Waterloo, Ontario, Ontario, NL2 3G1, Canada }

\author{Marek Korkusinski} 
\affiliation{National Research Council Canada, Ottawa, Ontario, K1A0R6, Canada }

\author{Dan  Dalacu} 
\affiliation{National Research Council Canada, Ottawa, Ontario, K1A0R6, Canada }

\author{Philip J. Poole} 
\affiliation{National Research Council Canada, Ottawa, Ontario, K1A0R6, Canada }

\author{Piotr Zawadzki} 
\affiliation{National Research Council Canada, Ottawa, Ontario, K1A0R6, Canada }

\author{Sergei Studenikin} 
\affiliation{National Research Council Canada, Ottawa, Ontario, K1A0R6, Canada }
\affiliation{Department of Physics and Astronomy, University of Waterloo, Waterloo, Ontario, Ontario, NL2 3G1, Canada }

 \author{Robin L. Williams} 
\affiliation{National Research Council Canada, Ottawa, Ontario, K1A0R6, Canada }
 
\author{Andrew S. Sachrajda} 
\affiliation{National Research Council Canada, Ottawa, Ontario, K1A0R6, Canada }

\author{Louis Gaudreau} 
\email[Corresponding author, email: ]{Louis.Gaudreau@nrc-cnrc.gc.ca} 
\affiliation{National Research Council Canada, Ottawa, Ontario, K1A0R6, Canada }
\date{\today}
 
\begin{abstract}

We study experimentally and theoretically the in-plane magnetic field dependence of the coupling between dots forming a vertically stacked double dot molecule. The InAsP molecule is grown epitaxially in an InP nanowire and interrogated optically at millikelvin temperatures. The strength of interdot tunneling, leading to the formation of the bonding-antibonding pair of molecular orbitals, is investigated by adjusting the sample geometry. For specific geometries, we show that the interdot coupling can be controlled in-situ using a magnetic field-mediated redistribution of interdot coupling strengths. This is an important milestone in the development of qubits required in future quantum information technologies.

\end{abstract}


\maketitle

Epitaxial quantum dots (QDs) embedded in nanowire waveguides are ideal sources of single and entangled photons due to the high collection efficiency and emission line purity which can be achieved with these devices \cite{ref:Gaussian,ref:Mirror,ref:Ultraclean,ref:DanReview}. Additionally, this architecture has the potential to form the building blocks of quantum information processors by coupling quantum dots in series within a nanowire. 
Quantum dot molecules with clear signatures of molecular bonding and antibonding states have been demonstrated, in which the carrier population can be tuned utilizing the quantum-confined Stark effect~\cite{ref:Ramanathan,ref:Stinaff}.
These optically active quantum dots are also very promising candidates for quantum networking units, because they can transfer the quantum information encoded in photonic qubits to solid state qubits and process this information in coupled quantum dot circuits~\cite{ref:Jennings,ref:Kim,ref:Gao}. 

Controlling the tunnel coupling between the dots is a key feature needed to appropriately tune and perform quantum gates between qubits. In electrostatically defined quantum dots, for example, interdot tunnel coupling can be achieved via electrical gates designed for this purpose and linear arrays of up to 9 qubits have been realized \cite{Zajac2016}. In epitaxial quantum dots, tunnel coupling is determined by the distance separating the quantum dots, which after the growth process can’t be changed \cite{ref:Jennings,ref:SA_VLS_tunable,ref:Froberg,ref:TelecomNW}. This creates reproducibility issues due to the uncertainties in the epitaxial growth on the atomistic level. Attempts to overcome these issues include measures aimed at introducing controlled structural changes, such as laser-induced intermixing \cite{ref:MeltingNW}, placing the emitters in a photonic cavity \cite{ref:TelecomNW}, or tuning strain fields in the vicinity of the dots \cite{ref:StrainTuning}. These processes result in an improved uniformity of the quantum dot emitters, however they do not allow for their time-dependent tuning and addressability. To enable this, external electric fields are applied to the dots by means of metallic gates, resulting in control of the charge states \cite{ref:GateTuning}, spectral tuning via the Stark shift \cite{ref:StarkShift} and control of the exciton fine structure via quadrupole fields \cite{ref:QuadrupoleTuning}. Additionally, recent electronic transport experiments in epitaxial quantum dots have demonstrated electrical tuning of tunnel coupling~\cite{Barker2019, Thomas2020, SadreMomtaz2020}. These approaches, however, require complex device design and engineering. In this letter, we demonstrate tunability of inter-dot coupling by applying a magnetic field perpendicular to the dot stacking direction. We first perform optical magnetospectroscopy of InAsP double quantum dots (DQDs) in InP nanowires and identify an inverse power law which governs the energy difference between the s-shell emission from each dot as a function of the inter-dot distance. Emission energies are affected by differences in dot composition and strain, while  coupling between dots is determined at the growth stage by the barrier thickness separating them. We will demonstrate, however, that we can tune, for specific states, the emission energy difference on-demand over a range of approximately 1\,meV by applying a magnetic field parallel to the plane of the quantum dots (i.e. Voigt geometry). As we will show, this energy shift is not possible without quantum-mechanical coupling between dots, and we interpret this result as an indication that magnetic field tuning of inter-dot tunnel coupling is occurring due to the quantum analogue of the classical Lorentz force.

The quantum dots used in this study are InAsP segments in a wurtzite crystal phase InP nanowire, grown using selective area vapour-liquid-solid epitaxy in a chemical beam epitaxy (CBE) system \cite{ref:Ultraclean,ref:SA_VLS}. The growth is catalysed by position and size controlled gold nanoparticles on an InP (111)B substrate. The size of the catalyst sets the diameter of the dot, $D$, and is defined using electron-beam lithography. Diameters between 20 and 22\,nm are used in this study. The InAsP segments are grown by injecting AsH$_3$ in the place of PH$_3$ into the CBE chamber for 3 seconds during the InP nanowire growth producing dots with thicknesses of $\sim$\,5\,nm. For double dot samples (Fig. 1a), different inter-dot spacers, $L$, are obtained in the same growth by virtue of a nanowire diameter-dependent growth rate \cite{ref:SA_VLS}.

\begin{figure}[htb!]
\centering
    \includegraphics[width=\linewidth]{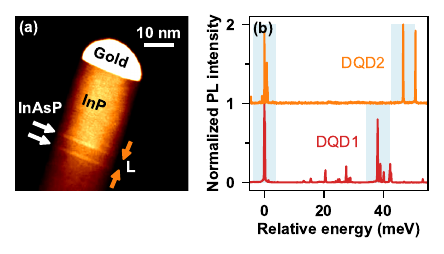}
\caption{\label{fig:DoubleDot}(a) Transmission electron microscopy image of a double dot nanowire where the spacer was grown for 5 seconds. White arrows indicate dots, orange arrows indicate inter-dot distance, $L$. (b) Normalized emission spectra of two double-dot wires for two different inter-dot distances. The low-energy s-shell peaks have been aligned to emphasize the relative energy of exciton emissions for these two cases. The s-shell X$^-$ states are highlighted by light blue boxes.}
\end{figure}

To enable efficient optical measurements the nanowires are further clad with a $\sim$\,100\,nm thick InP shell using growth conditions set to give a base diameter of $\sim 220$\,nm and a tip tapered to $\sim 20$\,nm. The thicker shell at the base supports the fundamental HE$_{11}$ waveguide mode that provides optimal coupling to the dot emission. The taper improves the coupling of the emitted light to the lensed fibre used in the spectroscopy measurements \cite{ref:Overcoming,ref:Gaussian}, i.e. the taper controls the numerical aperture of the source.

A direct measurement of the inter-dot separation, $L$, using electron microscopy techniques is not possible for clad nanowires. To determine $L$ we instead use the diameter-dependent non-linear growth rate model developed in Ref.~\citenum{ref:SA_VLS}. This approach only provides an estimate of the value of $L$ due to ($i$) a variation in nanowire diameters of $\pm 2$\,nm for nominally identical nanowires \cite{ref:SA_VLS_tunable},  ($ii$) a diameter-dependent incubation time to initiate the nanowire growth, and ($iii$) a reservoir effect which delays the onset of the spacer growth when switching from InAs to InP \cite{ref:Froberg}. We assume a 5-second delay to initiate the spacer growth and neglect the diameter-dependent incubation time. The 2 double dot samples studied (DQD1, DQD2) have dot diameters $D_1=20$\,nm and $D_2=22$\,nm. The first dot is incorporated into the InP nanowire after growing the base for 905 seconds and the second dot after growing a 15-second InP spacer on top of the first dot. Using the assumptions above, we estimate spacer thicknesses of $L_1 = 9.5\pm2$\,nm and $L_2 = 7\pm2$\,nm  for the two samples.

Our experiment uses a 780\,nm wavelength excitation laser coupled into the 1\% port of a 99-1 fibre beam splitter. The splitter's input port is connected to a 780HP single-mode fibre that runs inside a dilution refrigerator (base temperature of 10\,mK), where a lensed fibre focuses the light on the nanowire sample. The lensed fibre also collects the emission from the chosen individual nanowire, and the collected light is directed by the beam splitter to a spectrometer for analysis. X, Y, Z piezo position controllers are used to select each nanowire. The sample resides in the center of an 8\,T superconducting magnet in the Voigt configuration. Laser power levels at the sample range from 1\,nW to 1\,$\mu$W in our experiments.  See the Supplementary Material for more information on this setup \cite{ref:SupMat}.

The high crystal purity of our nanowires results in extremely narrow photoluminescence (PL) lines from the quantum dots, approaching the resolution of the spectrometer ($\sim$0.024\,nm). At the pumping levels used the samples exhibit emission spectra characterised by the lowest-energy (s-shell) exciton complexes: the neutral exciton (X), the negatively-charged exciton (X$^-$) and the neutral biexciton (XX). Emission lines are identified primarily through power-dependence measurements \cite{ref:AlGaAsSinglePhoton,ref:Mirror,ref:SolidStateSource,ref:Ultraclean} and analysis of their behaviour in a magnetic field~\cite{ref:SupMat}.  Tests were initially conducted on single-dot samples to observe the typical magnetic field dependence of emission lines associated with each s-shell exciton complex~\cite{ref:SupMat}. These single-dot studies were then used to establish a baseline for comparison with the double dots to determine if they are coupled or not. Double-dot samples which display behaviour similar to single dot structures can be assumed to have no inter-dot coupling, while observed differences are assumed to arise from interactions between dots.

Figure \ref{fig:DoubleDot}(b) shows the s-shell emission spectra of two separate DQD samples with different inter-dot distances, plotted relative to the lowest s-shell X$^-$ peak. Two distinct sets of peaks are observed, each associated with one of the quantum dots, and we refer to each set as either as the `low-energy' or `high-energy' dot based on their relative s-shell emission energies. The figure shows an increase in the energy difference between the ground state emission of the two quantum dots (indicated by the light square boxes in each trace) as $L$ is decreased. There are two mechanisms that result in a splitting of the emission energies: a difference in the physical properties of the dots such as composition and size, and a coupling between the dots. In this nanowire system, we typically observe an energy difference between nominally identical dots even for dots separated by $L>100$\,nm where no coupling is expected. This non-degeneracy is associated with a difference in As content between the two dots\cite{ref:TelecomNW} and in the limit of large $L$, we expect the emission energy difference to be solely due to this difference in As content. With decreasing $L$ we expect an increasing contribution due to tunnel coupling.

In coupled systems\cite{ref:DQDSplitting}, the strengthening of the inter-dot coupling (the increase in energy difference between the s-shell energies of the two dots, $\Delta E_{S}$) can be described phenomenologically using a 1/$L^3$ dependence\cite{ref:DoubleDots,ref:SolidStateSource}. For our data we obtain: \begin{equation}\label{eq:LtoNeg3}
\Delta E_{S} = \frac{33.0\times10^3 \text{ meV}\cdot\text{nm$^3$}}{(L+4.88 \text{ nm})^3}+27.0 \text{ meV}
\end{equation} Since $L$ is measured from the edge of each dot, the offset in inter-dot separation (4.88 nm) is a result of the finite thickness of the dots, which is roughly 4.5 nm. The energy offset of 27 meV is included to account for the difference in composition between the two dots within a nanowire which produces an additional splitting as discussed above.

\begin{figure}
\includegraphics[width=\linewidth]{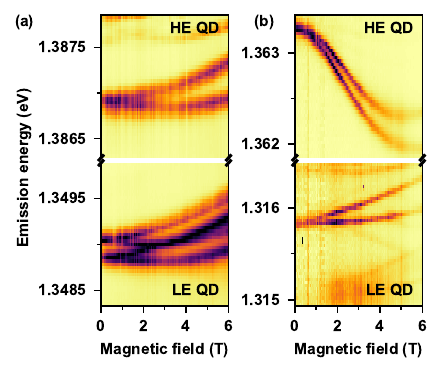}
\caption{\label{fig:DDM}Magnetic field dependence of the emission spectra of (a) DQD1 with inter-dot spacing of $L_1=9.5$\,nm and (b) DQD2 with $L_2=7$\,nm. Magnetic decoupling of the high-energy (HE) and low-energy (LE) dots is observable in (b) as a downward shift in the energy of the HE state.}
\end{figure}

These data alone are not conclusive proof that inter-dot coupling is responsible for the majority of the detuning as dot separation decreases. We therefore provide more direct evidence of quantum mechanical coupling through magneto-PL studies in the Voigt configuration. The magneto-PL of double dot samples reveals the combined effects of the field strength and inter-dot separation on the PL spectrum. Figure \ref{fig:DDM} compares the PL spectra of the high-energy and low-energy dots as a function of magnetic field for samples DQD1 and DQD2 with inter-dot distances of $L_1=9.5$\,nm (Fig. \ref{fig:DDM}a) and $L_2=7$\,nm (Fig. \ref{fig:DDM}b), respectively. DQD1 ($L_1=9.5$\,nm) exhibits both Zeeman splitting and diamagnetic shifts, and the spectrum of each dot is qualitatively similar to that of a single-dot sample (see Supplementary Material \cite{ref:SupMat}). The large separation between the dots reduces the tunnel coupling between them to the degree that they effectively behave as independent systems.

\begin{figure*}
\includegraphics[width = \linewidth]{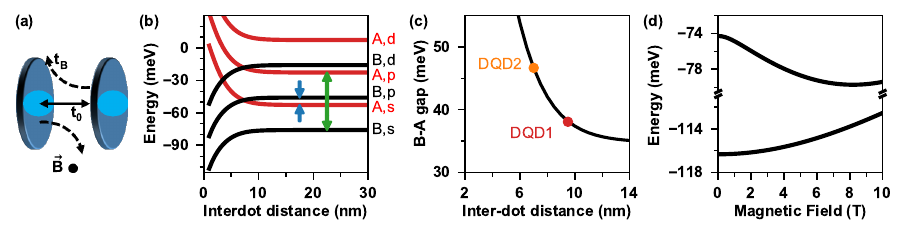}
\caption{\label{fig:Lorentz}(a) Schematic representation of the tunnelling direction of electrons between dots for zero magnetic field, $t_0$ (solid line), and high field, $t_B$ (dashed lines), in the Voigt configuration. Light blue shading indicates the modified confinement region induced by the B-field. (b) Bonding (black) and antibonding (red) quantum molecular states of an electron. The blue arrows indicate the states whose coupling generates the emission line configuration in panel (d) and Fig.~\ref{fig:DDM}(b). (c) The modelled energy gap between s-shell emission lines of the high- and low-energy dots as a function of inter-dot distance, $L$. The energy gaps from Fig.~\ref{fig:DoubleDot}(b) are overlaid on the model curve, allowing us to verify the values of $L$ for those samples. (d) Emission line model showing the same shift in tunnel coupling strength as observed in Fig.~\ref{fig:DoubleDot}(b), without Zeeman splitting.}
\end{figure*}

Similarly, the low-energy QD in DQD2 ($L_2=7$\,nm) shows a magnetic field dependence similar to that of a single-dot nanowire. In contrast, the high-energy QD in this sample reveals a downward shift in energy with increasing B-field, a clear signature of a modification of the interdot coupling, which is the main finding of this work. We interpret this energy shift as a quantum analogue of the classical Lorentz force inducing an effective decoupling between the QDs. In the Voigt configuration the magnetic field is applied perpendicularly to the tunneling direction, and as carriers tunnel between dots, the Lorentz force alters their trajectories, leading to a redistribution of inter-dot coupling strengths. This concept is schematically shown in Fig. \ref{fig:Lorentz}(a). The decoupling mechanism reduces the energy difference between the emissions of the QDs from 47.5\,meV  to 46\,meV which causes an effective increase in inter-dot distance from $L_2=7$\,nm to 10\,nm.

We now perform a more quantitative analysis of the emission spectra of the coupled quantum dot molecule by approximating the double-dot confinement with a simple, separable three-dimensional double-well potential~\cite{ref:SupMat}. The Hamiltonian of the system is written as $\hat{H}=\hat{T}+\hat{V}_{l}(x,y)+\hat{V}_v(z)$, with $\hat{T}=\hat{p}^2/2m$ being the kinetic energy operator of the particle with mass $m$. The lateral confinement $\hat{V}_{l}(x,y)$ of either dot is approximated by a two-dimensional parabolic potential with frequency $\hbar\Omega_e=30$\,meV for the electron, and $\hbar\Omega_h=15$\,meV for the hole \cite{ref:Marek1}. Along the nanowire axis we deal with a one-dimensional double-well potential $\hat{V}_v(z)$ with equal well widths (dot heights) $H=4.5$\,nm (effectively accounting for interface effects). The difference in dot composition, however, means the confinement depths are such that $V_1 > V_2$, where 1 (2) corresponds to the low (high) energy QD. We calculate the single-particle energies and wave functions in this 3D confinement numerically. Utilizing the effective masses for the electron, $m_e=0.03$ $m_0$, and the hole, $m_h=0.06$ $m_0$, and assuming an exciton binding energy of $25$\,meV \cite{ref:Marek1}, we can find the well depths $V_1$ and $V_2$ by fitting the model positions of the emission maxima from either dot to the measured results of uncoupled dots, as in Fig.~\ref{fig:DDM}(a) at the magnetic field $B=0$ T. We obtain barrier height potentials $V_1^e=239$\,meV and $V_2^e=203$\,meV for the electron, and $V_1^h=119.5$\,meV and $V_2^h=101.5$\,meV for the hole, counting from the band edges of the InP barrier material.

With the model fully parametrized, we now compute the energy levels of the confined carriers as a function of the inter-dot distance. In Fig.~\ref{fig:Lorentz}(b) we show the bonding (black lines) and antibonding (red lines) states of an electron, including the three lowest shells, originating from the harmonic lateral confinement. We detect a significant tunnel coupling for inter-dot distances below $10$\,nm, and a rich structure of level crossings arising from the opposite shifts in levels of different vertical symmetry (bonding or antibonding). Further, by performing the calculation for both types of carriers, we may plot the energy difference between the emission maxima of excitons built on the bonding and antibonding quantum molecular states (Fig.~\ref{fig:Lorentz}c). By taking the experimentally measured energy gaps and placing them on the modelled curve in Fig.~\ref{fig:Lorentz}(c) we determine interdot distance of $9.5$\,nm and $7$\,nm, corresponding to the samples DQD1 and DQD2, as marked in the figure. The model curve in Fig.~\ref{fig:Lorentz}(c) is in agreement with our empirical fit of the data to a $1/L^3$ relation (Eq.~1), for which we used these values of $L$.

We then include the effects of the magnetic field in the Voigt configuration into our model by introducing the generalized momentum operator for the electron (hole) $\hat{\vec{\pi}}=\hat{\vec{p}}\pm \frac{e}{c}\vec{A}$, where $e$ is the unit charge, $c$ is the speed of light, and $\vec{A}$ is the magnetic vector potential \cite{ref:SupMat}. With the choice $\vec{A}=B[0,0,y]$, $B$ being the magnetic field magnitude, the kinetic energy expressed by this momentum acquires two new terms: ($i$) a harmonic correction to the lateral confinement in the $y$ direction, responsible for the diamagnetic shift of energy levels, and ($ii$) an electron (hole) Hamiltonian term $\hat{H}_{yz}=\mp i\hbar\Omega_c y \partial/\partial z$, where $\Omega_c=eB/mc$ is the cyclotron energy. We include the two new terms in the basis of the zero-field bonding and antibonding states. Owing to its bilinear form, the term $\hat{H}_{yz}$ couples the vertical and horizontal wave function components with opposite symmetry. The relevant coupling scheme is marked in Fig.~\ref{fig:Lorentz}(b) with arrows. We find that the bonding $s$-shell orbital $B,s$ is coupled to one of the antibonding $p$-shell orbitals $A,p$ (as marked by the green double-arrow). On the other hand, the antibonding $s$-shell orbital $A,s$ couples with the bonding $p$-shell orbital $B,p$ (as marked by the blue arrows). The strength of this coupling depends on the magnetic field, but also on the energy gap between the relevant levels (large gap for the former pair, small gap for the latter). As a result, the lowest-energy antibonding level, $A,s$, will experience a strong shift towards the lower energies for inter-dot distances larger than $\sim 7$ nm (i.e., past the crossing of the $A,s$ and $B,p$ levels), but the level $B,s$ will not experience such a shift. This is evident in Fig.~\ref{fig:Lorentz}(d), where we plot the calculated energies of the two lowest orbitals as a function of the magnetic field. The lower-energy orbital experiences only a diamagnetic shift, while the higher-energy excited state exhibits both a downward shift arising from $\hat{H}_{yz}$ and an upward diamagnetic shift. As these two levels form the two optically active excitonic states, this non-monotonic shift is precisely what is seen in the experimental spectra of Fig.~\ref{fig:DDM}(b).

We stress that the magnetic coupling mechanism does not renormalize the zero-field tunneling elements directly. Indeed, these elements are relevant in the creation of the bonding and antibonding orbitals of the same lateral symmetry (e.g., $B,s$ and $A,s$). However, the appearance of the characteristic shift as a function of the magnetic field  guarantees that there is an overlap between the levels of the two dots, and therefore constitutes a spectroscopic proof of the inter-dot quantum-mechanical coupling. The decrease of the energy distance between the peaks can be perceived as an effective renormalization of the bonding-antibonding gap, however it is a result of a complex interaction of various levels.

In summary, we have studied the magnetospectroscopy of epitaxial InAsP quantum dots, demonstrating an inverse power relationship between the double-dot energy difference and the inter-dot separation distance. Additionally, we have provided conclusive proof of a technique for tuning the inter-dot tunnel coupling. This technique does not require the fabrication of a series of samples, but it instead relies on the magnetic field's influence on the electron and hole wave functions. In the sample DQD2 with $L_2=7$\,nm, we were able to achieve in-situ tuning of the tunnel coupling through the application of a magnetic field in the Voigt configuration, as demonstrated by a 1\,meV reduction in emission energy detuning. Future experiments may involve studies of the level structure of the dots by populating them with more carriers, allowing access to higher excited states. The combination of the Lorentz effect and the Voigt configuration have previously been used to map out the wave functions of electrons confined in dots \cite{ref:Map1,ref:Map2}, which we could now observe as a function of the coupling strength. The small variations in dot composition have important implications for experimental reproducibility \cite{ref:Marek1}, so the difference in As content between dots should be addressed through further refinements in the growth process. Overall, our experimental results may have potential applications in quantum information technologies based on epitaxial quantum dot qubits, especially where two-qubit gates need to be performed and precise adjustment of the tunnel coupling is required. Even though it leads to substantial shifts of the positions of spectral lines, the application of magnetic fields may be too slow to perform
efficient quantum gates. However, the coarse magnetic tuning of the tunnel coupling can be supplemented by finer adjustments carried out by electric gates, which can operate at high frequencies~\cite{SadreMomtaz2020,ReimerNL,Fasth}. Moreover, these nanowire DQDs can be used to generate entangled photon pairs for quantum communication, and our results suggest a technique for adjusting the degree of entanglement by tuning the inter-dot coupling.

J.P. and S. S. thank the Natural Sciences and Engineering Research Council of Canada for financial support.

\bibliography{References}

\providecommand{\latin}[1]{#1}
\makeatletter
\providecommand{\doi}
  {\begingroup\let\do\@makeother\dospecials
  \catcode`\{=1 \catcode`\}=2 \doi@aux}
\providecommand{\doi@aux}[1]{\endgroup\texttt{#1}}
\makeatother
\providecommand*\mcitethebibliography{\thebibliography}
\csname @ifundefined\endcsname{endmcitethebibliography}
  {\let\endmcitethebibliography\endthebibliography}{}
\begin{mcitethebibliography}{34}
\providecommand*\natexlab[1]{#1}
\providecommand*\mciteSetBstSublistMode[1]{}
\providecommand*\mciteSetBstMaxWidthForm[2]{}
\providecommand*\mciteBstWouldAddEndPuncttrue
  {\def\EndOfBibitem{\unskip.}}
\providecommand*\mciteBstWouldAddEndPunctfalse
  {\let\EndOfBibitem\relax}
\providecommand*\mciteSetBstMidEndSepPunct[3]{}
\providecommand*\mciteSetBstSublistLabelBeginEnd[3]{}
\providecommand*\EndOfBibitem{}
\mciteSetBstSublistMode{f}
\mciteSetBstMaxWidthForm{subitem}{(\alph{mcitesubitemcount})}
\mciteSetBstSublistLabelBeginEnd
  {\mcitemaxwidthsubitemform\space}
  {\relax}
  {\relax}

\bibitem[Bulgarini \latin{et~al.}(2014)Bulgarini, Reimer, Bavinck, J{\"o}ns,
  Dalacu, Poole, Bakkers, and Zwiller]{ref:Gaussian}
Bulgarini,~G.; Reimer,~M.~E.; Bavinck,~M.~B.; J{\"o}ns,~K.~D.; Dalacu,~D.;
  Poole,~P.~J.; Bakkers,~E. P. A.~M.; Zwiller,~V. Nanowire Waveguides Launching
  Single Photons in a {G}aussian Mode for Ideal Fiber Coupling. \emph{Nano
  Lett.} \textbf{2014}, \emph{14}, 4102\relax
\mciteBstWouldAddEndPuncttrue
\mciteSetBstMidEndSepPunct{\mcitedefaultmidpunct}
{\mcitedefaultendpunct}{\mcitedefaultseppunct}\relax
\EndOfBibitem
\bibitem[Reimer \latin{et~al.}(2012)Reimer, Bulgarini, Akopian, Hocevar,
  Bavinck, Verheijen, Bakkers, Kouwenhoven, and Zwiller]{ref:Mirror}
Reimer,~M.~E.; Bulgarini,~G.; Akopian,~N.; Hocevar,~M.; Bavinck,~M.~B.;
  Verheijen,~M.~A.; Bakkers,~E. P. A.~M.; Kouwenhoven,~L.~P.; Zwiller,~V.
  Bright single-photon sources in bottom-up tailored nanowires. \emph{Nat.
  Commun.} \textbf{2012}, \emph{3}, 737\relax
\mciteBstWouldAddEndPuncttrue
\mciteSetBstMidEndSepPunct{\mcitedefaultmidpunct}
{\mcitedefaultendpunct}{\mcitedefaultseppunct}\relax
\EndOfBibitem
\bibitem[Dalacu \latin{et~al.}(2012)Dalacu, Mnaymneh, Lapointe, Wu, Poole,
  Bulgarini, Zwiller, and Reimer]{ref:Ultraclean}
Dalacu,~D.; Mnaymneh,~K.; Lapointe,~J.; Wu,~X.; Poole,~P.~J.; Bulgarini,~G.;
  Zwiller,~V.; Reimer,~M.~E. Ultraclean Emission from {I}n{A}s{P} Quantum Dots
  in Defect-Free Wurtzite {I}n{P} Nanowires. \emph{Nano Lett.} \textbf{2012},
  \emph{12}, 5919\relax
\mciteBstWouldAddEndPuncttrue
\mciteSetBstMidEndSepPunct{\mcitedefaultmidpunct}
{\mcitedefaultendpunct}{\mcitedefaultseppunct}\relax
\EndOfBibitem
\bibitem[Dalacu \latin{et~al.}(2019)Dalacu, Poole, and Williams]{ref:DanReview}
Dalacu,~D.; Poole,~P.~J.; Williams,~R.~L. Nanowire-based sources of
  non-classical light. \emph{Nanotechnology} \textbf{2019}, \emph{30},
  232001\relax
\mciteBstWouldAddEndPuncttrue
\mciteSetBstMidEndSepPunct{\mcitedefaultmidpunct}
{\mcitedefaultendpunct}{\mcitedefaultseppunct}\relax
\EndOfBibitem
\bibitem[Ramanathan \latin{et~al.}(2013)Ramanathan, Petersen, Wijesundara,
  Thota, Stinaff, Kerfoot, Scheibner, Bracker, and Gammon]{ref:Ramanathan}
Ramanathan,~S.; Petersen,~G.; Wijesundara,~K.; Thota,~R.; Stinaff,~E.;
  Kerfoot,~M.; Scheibner,~M.; Bracker,~A.; Gammon,~D. Quantum-confined Stark
  effects in coupled {I}n{A}s/{G}a{A}s quantum dots. \emph{Appl. Phys. Lett.}
  \textbf{2013}, \emph{102}, 213101\relax
\mciteBstWouldAddEndPuncttrue
\mciteSetBstMidEndSepPunct{\mcitedefaultmidpunct}
{\mcitedefaultendpunct}{\mcitedefaultseppunct}\relax
\EndOfBibitem
\bibitem[Stinaff \latin{et~al.}(2006)Stinaff, Scheibner, Bracker, Ponomarev,
  Korenev, Ware, Doty, Reinecke, and Gammon]{ref:Stinaff}
Stinaff,~E.~A.; Scheibner,~M.; Bracker,~A.~S.; Ponomarev,~I.~V.;
  Korenev,~V.~L.; Ware,~M.~E.; Doty,~M.~F.; Reinecke,~T.~L.; Gammon,~D. Optical
  Signatures of Coupled Quantum Dots. \emph{Science} \textbf{2006}, \emph{311},
  636\relax
\mciteBstWouldAddEndPuncttrue
\mciteSetBstMidEndSepPunct{\mcitedefaultmidpunct}
{\mcitedefaultendpunct}{\mcitedefaultseppunct}\relax
\EndOfBibitem
\bibitem[Jennings \latin{et~al.}(2019)Jennings, Ma, Wickramasinghe, Doty,
  Scheibner, Stinaff, and Ware]{ref:Jennings}
Jennings,~C.; Ma,~X.; Wickramasinghe,~T.; Doty,~M.; Scheibner,~M.; Stinaff,~E.;
  Ware,~M. Self-Assembled {I}n{A}s/{G}a{A}s Coupled Quantum Dots for Qhotonic
  Quantum Technologies. \emph{Adv. Quantum Tech.} \textbf{2019}, \emph{3},
  1900085\relax
\mciteBstWouldAddEndPuncttrue
\mciteSetBstMidEndSepPunct{\mcitedefaultmidpunct}
{\mcitedefaultendpunct}{\mcitedefaultseppunct}\relax
\EndOfBibitem
\bibitem[Kim \latin{et~al.}(2011)Kim, Carter, Greilich, Bracker, and
  Gammon]{ref:Kim}
Kim,~D.; Carter,~S.~G.; Greilich,~A.; Bracker,~A.~S.; Gammon,~D. Ultrafast
  optical control of entanglement between two quantum-dot spins. \emph{Nature
  Phys.} \textbf{2011}, \emph{7}, 223\relax
\mciteBstWouldAddEndPuncttrue
\mciteSetBstMidEndSepPunct{\mcitedefaultmidpunct}
{\mcitedefaultendpunct}{\mcitedefaultseppunct}\relax
\EndOfBibitem
\bibitem[Gao \latin{et~al.}(2012)Gao, Fallahi, Togan, Miguel-Sanchez, and
  Imamoglu]{ref:Gao}
Gao,~W.~B.; Fallahi,~P.; Togan,~E.; Miguel-Sanchez,~J.; Imamoglu,~A.
  Observation of entanglement between a quantum dot spin and a single photon.
  \emph{Nature} \textbf{2012}, \emph{491}, 426\relax
\mciteBstWouldAddEndPuncttrue
\mciteSetBstMidEndSepPunct{\mcitedefaultmidpunct}
{\mcitedefaultendpunct}{\mcitedefaultseppunct}\relax
\EndOfBibitem
\bibitem[Zajac \latin{et~al.}(2016)Zajac, Hazard, Mi, Nielsen, and
  Petta]{Zajac2016}
Zajac,~D.~M.; Hazard,~T.~M.; Mi,~X.; Nielsen,~E.; Petta,~J.~R. {Scalable Gate
  Architecture for a One-Dimensional Array of Semiconductor Spin Qubits}.
  \emph{Physical Review Applied} \textbf{2016}, \emph{6}, 054013\relax
\mciteBstWouldAddEndPuncttrue
\mciteSetBstMidEndSepPunct{\mcitedefaultmidpunct}
{\mcitedefaultendpunct}{\mcitedefaultseppunct}\relax
\EndOfBibitem
\bibitem[Dalacu \latin{et~al.}(2011)Dalacu, Mnaymneh, Wu, Lapointe, Aers,
  Poole, and Williams]{ref:SA_VLS_tunable}
Dalacu,~D.; Mnaymneh,~K.; Wu,~X.; Lapointe,~J.; Aers,~G.~C.; Poole,~P.~J.;
  Williams,~R.~L. Selective-area vapor-liquid-solid growth of tunable
  {I}n{A}s{P} quantum dots in nanowires. \emph{Appl. Phys. Lett.}
  \textbf{2011}, \emph{98}, 251101\relax
\mciteBstWouldAddEndPuncttrue
\mciteSetBstMidEndSepPunct{\mcitedefaultmidpunct}
{\mcitedefaultendpunct}{\mcitedefaultseppunct}\relax
\EndOfBibitem
\bibitem[Fr{\"o}berg \latin{et~al.}(2008)Fr{\"o}berg, Wacaser, Wagner,
  Jeppesen, Ohlsson, Deppert, and Samuelson]{ref:Froberg}
Fr{\"o}berg,~L.~E.; Wacaser,~B.~A.; Wagner,~J.~B.; Jeppesen,~S.;
  Ohlsson,~B.~J.; Deppert,~K.; Samuelson,~L. Transients in the formation of
  nanowire heterostructures. \emph{Nano Lett.} \textbf{2008}, \emph{8},
  3815\relax
\mciteBstWouldAddEndPuncttrue
\mciteSetBstMidEndSepPunct{\mcitedefaultmidpunct}
{\mcitedefaultendpunct}{\mcitedefaultseppunct}\relax
\EndOfBibitem
\bibitem[Haffouz \latin{et~al.}(2018)Haffouz, Zeuner, Dalacu, Poole, Lapointe,
  Poitras, Mnaymneh, X.~Wu, Korkusinski, Sch{\"o}ll, J{\"o}ns, Zwiller, and
  Williams]{ref:TelecomNW}
Haffouz,~S.; Zeuner,~K.~D.; Dalacu,~D.; Poole,~P.~J.; Lapointe,~J.;
  Poitras,~D.; Mnaymneh,~K.; X.~Wu,~M.~C.; Korkusinski,~M.; Sch{\"o}ll,~E.;
  J{\"o}ns,~K.~D.; Zwiller,~V.; Williams,~R.~L. Bright Single {I}n{A}s{P}
  Quantum Dots at Telecom Wavelengths in Position-Controlled {I}n{P} Nanowires:
  The Role of the Photonic Waveguide. \emph{Nano Lett.} \textbf{2018},
  \emph{18}, 3047\relax
\mciteBstWouldAddEndPuncttrue
\mciteSetBstMidEndSepPunct{\mcitedefaultmidpunct}
{\mcitedefaultendpunct}{\mcitedefaultseppunct}\relax
\EndOfBibitem
\bibitem[Fiset-Cyr \latin{et~al.}(2018)Fiset-Cyr, Dalacu, Haffouz, Poole,
  Lapointe, Aers, and Williams]{ref:MeltingNW}
Fiset-Cyr,~A.; Dalacu,~D.; Haffouz,~S.; Poole,~P.~J.; Lapointe,~J.;
  Aers,~G.~C.; Williams,~R.~L. In-situ tuning of individual position-controlled
  nanowire quantum dots via laser-induced intermixing. \emph{Appl. Phys. Lett.}
  \textbf{2018}, \emph{113}, 053105\relax
\mciteBstWouldAddEndPuncttrue
\mciteSetBstMidEndSepPunct{\mcitedefaultmidpunct}
{\mcitedefaultendpunct}{\mcitedefaultseppunct}\relax
\EndOfBibitem
\bibitem[Grim \latin{et~al.}(2019)Grim, Bracker, Zalalutdinov, Carter, Kozen,
  Kim, Kim, Mlack, Yakes, Lee, and Gammon]{ref:StrainTuning}
Grim,~J.; Bracker,~A.; Zalalutdinov,~M.; Carter,~S.; Kozen,~A.; Kim,~M.;
  Kim,~C.; Mlack,~J.; Yakes,~M.; Lee,~B.; Gammon,~D. Scalable in operando
  strain tuning in nanophotonic waveguides enabling three-quantum-dot
  superradiance. \emph{Nat. Mater.} \textbf{2019}, \emph{18}, 963\relax
\mciteBstWouldAddEndPuncttrue
\mciteSetBstMidEndSepPunct{\mcitedefaultmidpunct}
{\mcitedefaultendpunct}{\mcitedefaultseppunct}\relax
\EndOfBibitem
\bibitem[Yang \latin{et~al.}(2016)Yang, Carter, Bracker, Yakes, Kim, Kim, Vora,
  and Gammon]{ref:GateTuning}
Yang,~L.; Carter,~S.; Bracker,~A.; Yakes,~M.; Kim,~M.; Kim,~C.; Vora,~P.;
  Gammon,~D. Optical spectroscopy of site-controlled quantum dots in a
  {S}chottky diode. \emph{Appl. Phys. Lett.} \textbf{2016}, \emph{108},
  233102\relax
\mciteBstWouldAddEndPuncttrue
\mciteSetBstMidEndSepPunct{\mcitedefaultmidpunct}
{\mcitedefaultendpunct}{\mcitedefaultseppunct}\relax
\EndOfBibitem
\bibitem[Ramanathan \latin{et~al.}(2013)Ramanathan, Petersen, Wijesundara,
  Thota, Stinaff, Kerfoot, Scheibner, Bracker, and Gammon]{ref:StarkShift}
Ramanathan,~S.; Petersen,~G.; Wijesundara,~K.; Thota,~R.; Stinaff,~E.;
  Kerfoot,~M.; Scheibner,~M.; Bracker,~A.; Gammon,~D. Quantum-confined {S}tark
  effects in coupled {I}n{A}s/{G}a{A}s quantum dots. \emph{Appl. Phys. Lett.}
  \textbf{2013}, \emph{102}, 213101\relax
\mciteBstWouldAddEndPuncttrue
\mciteSetBstMidEndSepPunct{\mcitedefaultmidpunct}
{\mcitedefaultendpunct}{\mcitedefaultseppunct}\relax
\EndOfBibitem
\bibitem[Zeeshan \latin{et~al.}(2019)Zeeshan, Sherlekar, Ahmadi, Williams, and
  Reimer]{ref:QuadrupoleTuning}
Zeeshan,~M.; Sherlekar,~N.; Ahmadi,~A.; Williams,~R.; Reimer,~M. Proposed
  scheme to generate bright entangled photon pairs by application of a
  quadrupole field to a single quantum dot. \emph{Phys. Rev. Lett.}
  \textbf{2019}, \emph{122}, 227401\relax
\mciteBstWouldAddEndPuncttrue
\mciteSetBstMidEndSepPunct{\mcitedefaultmidpunct}
{\mcitedefaultendpunct}{\mcitedefaultseppunct}\relax
\EndOfBibitem
\bibitem[Barker \latin{et~al.}(2019)Barker, Lehmann, Namazi, Nilsson,
  Thelander, Dick, and Maisi]{Barker2019}
Barker,~D.; Lehmann,~S.; Namazi,~L.; Nilsson,~M.; Thelander,~C.; Dick,~K.~A.;
  Maisi,~V.~F. {Individually addressable double quantum dots formed with
  nanowire polytypes and identified by epitaxial markers}. \emph{Appl. Phys.
  Lett.} \textbf{2019}, \emph{114}, 183502\relax
\mciteBstWouldAddEndPuncttrue
\mciteSetBstMidEndSepPunct{\mcitedefaultmidpunct}
{\mcitedefaultendpunct}{\mcitedefaultseppunct}\relax
\EndOfBibitem
\bibitem[Thomas \latin{et~al.}(2020)Thomas, Baumgartner, Gubser, J{\"{u}}nger,
  F{\"{u}}l{\"{o}}p, Nilsson, Rossi, Zannier, Sorba, and
  Sch{\"{o}}nenberger]{Thomas2020}
Thomas,~F.~S.; Baumgartner,~A.; Gubser,~L.; J{\"{u}}nger,~C.;
  F{\"{u}}l{\"{o}}p,~G.; Nilsson,~M.; Rossi,~F.; Zannier,~V.; Sorba,~L.;
  Sch{\"{o}}nenberger,~C. {Highly symmetric and tunable tunnel couplings in
  InAs/InP nanowire heterostructure quantum dots}. \emph{Nanotechnology}
  \textbf{2020}, \emph{31}, 135003\relax
\mciteBstWouldAddEndPuncttrue
\mciteSetBstMidEndSepPunct{\mcitedefaultmidpunct}
{\mcitedefaultendpunct}{\mcitedefaultseppunct}\relax
\EndOfBibitem
\bibitem[{Sadre Momtaz} \latin{et~al.}(2020){Sadre Momtaz}, Servino, Demontis,
  Zannier, Ercolani, Rossi, Rossella, Sorba, Beltram, and
  Roddaro]{SadreMomtaz2020}
{Sadre Momtaz},~Z.; Servino,~S.; Demontis,~V.; Zannier,~V.; Ercolani,~D.;
  Rossi,~F.; Rossella,~F.; Sorba,~L.; Beltram,~F.; Roddaro,~S. {Orbital Tuning
  of Tunnel Coupling in InAs/InP Nanowire Quantum Dots}. \emph{Nano Letters}
  \textbf{2020}, \emph{20}, 1693\relax
\mciteBstWouldAddEndPuncttrue
\mciteSetBstMidEndSepPunct{\mcitedefaultmidpunct}
{\mcitedefaultendpunct}{\mcitedefaultseppunct}\relax
\EndOfBibitem
\bibitem[Dalacu \latin{et~al.}(2009)Dalacu, Kam, Austing, Wu, Lapointe, Aers,
  and Poole]{ref:SA_VLS}
Dalacu,~D.; Kam,~A.; Austing,~D.~G.; Wu,~X.; Lapointe,~J.; Aers,~G.~C.;
  Poole,~P.~J. Selective-area vapour–liquid–solid growth of {I}n{P}
  nanowires. \emph{Nanotechnology} \textbf{2009}, \emph{20}, 395602\relax
\mciteBstWouldAddEndPuncttrue
\mciteSetBstMidEndSepPunct{\mcitedefaultmidpunct}
{\mcitedefaultendpunct}{\mcitedefaultseppunct}\relax
\EndOfBibitem
\bibitem[Reimer \latin{et~al.}(2016)Reimer, Bulgarini, Fognini, Heeres, Witek,
  Versteegh, Rubino, Braun, Kamp, H{\"o}fling, Dalacu, Lapointe, Poole, and
  Zwiller]{ref:Overcoming}
Reimer,~M.~E.; Bulgarini,~G.; Fognini,~A.; Heeres,~R.~W.; Witek,~B.~J.;
  Versteegh,~M. A.~M.; Rubino,~A.; Braun,~T.; Kamp,~M.; H{\"o}fling,~S.;
  Dalacu,~D.; Lapointe,~J.; Poole,~P.~J.; Zwiller,~V. Overcoming power
  broadening of the quantum dot emission in a pure wurtzite nanowire.
  \emph{Phys. Rev. B} \textbf{2016}, \emph{93}, 195316\relax
\mciteBstWouldAddEndPuncttrue
\mciteSetBstMidEndSepPunct{\mcitedefaultmidpunct}
{\mcitedefaultendpunct}{\mcitedefaultseppunct}\relax
\EndOfBibitem
\bibitem[ref()]{ref:SupMat}
See {S}upplemental {M}aterial at [{URL}] for further details on the
  experimental setup, the single-dot experiments, and the {DQD} coupling
  model.\relax
\mciteBstWouldAddEndPunctfalse
\mciteSetBstMidEndSepPunct{\mcitedefaultmidpunct}
{}{\mcitedefaultseppunct}\relax
\EndOfBibitem
\bibitem[Heinrich \latin{et~al.}(2010)Heinrich, Huggenberger, Heindel,
  Reitzenstein, H{\"o}fling, Worschech, and Forchel]{ref:AlGaAsSinglePhoton}
Heinrich,~J.; Huggenberger,~A.; Heindel,~T.; Reitzenstein,~S.; H{\"o}fling,~S.;
  Worschech,~L.; Forchel,~A. Single photon emission from positioned
  {G}a{A}s/{A}l{G}a{A}s photonic nanowires. \emph{Appl. Phys. Lett.}
  \textbf{2010}, \emph{96}, 211117\relax
\mciteBstWouldAddEndPuncttrue
\mciteSetBstMidEndSepPunct{\mcitedefaultmidpunct}
{\mcitedefaultendpunct}{\mcitedefaultseppunct}\relax
\EndOfBibitem
\bibitem[Khoshnegar \latin{et~al.}(2017)Khoshnegar, Huber, Predojevi{\'c},
  Dalacu, Prilm{\"u}ller, Lapointe, Wu, Tamarat, Lounis, Poole, Weihs, and
  Majedi]{ref:SolidStateSource}
Khoshnegar,~M.; Huber,~T.; Predojevi{\'c},~A.; Dalacu,~D.; Prilm{\"u}ller,~M.;
  Lapointe,~J.; Wu,~X.; Tamarat,~P.; Lounis,~B.; Poole,~P.; Weihs,~G.;
  Majedi,~H. A solid state source of photon triplets based on quantum dot
  molecules. \emph{Nat. Commun.} \textbf{2017}, \emph{8}, 15716\relax
\mciteBstWouldAddEndPuncttrue
\mciteSetBstMidEndSepPunct{\mcitedefaultmidpunct}
{\mcitedefaultendpunct}{\mcitedefaultseppunct}\relax
\EndOfBibitem
\bibitem[Bayer \latin{et~al.}(2000)Bayer, Hawrylak, Hinzer, Fafard,
  Korkusinski, Wasilewski, Stern, and Forchel]{ref:DQDSplitting}
Bayer,~M.; Hawrylak,~P.; Hinzer,~K.; Fafard,~S.; Korkusinski,~M.;
  Wasilewski,~Z.~R.; Stern,~O.; Forchel,~A. Coupling and Entangling of Quantum
  States in Quantum Dot Molecules. \emph{Science} \textbf{2000}, \emph{291},
  451\relax
\mciteBstWouldAddEndPuncttrue
\mciteSetBstMidEndSepPunct{\mcitedefaultmidpunct}
{\mcitedefaultendpunct}{\mcitedefaultseppunct}\relax
\EndOfBibitem
\bibitem[Carlson \latin{et~al.}(2019)Carlson, Dalacu, Gustin, Haffouz, Wu,
  Lapointe, Williams, Poole, and Hughes]{ref:DoubleDots}
Carlson,~C.; Dalacu,~D.; Gustin,~C.; Haffouz,~S.; Wu,~X.; Lapointe,~J.;
  Williams,~R.~L.; Poole,~P.~J.; Hughes,~S. Theory and experiments of coherent
  photon coupling in semiconductor nanowire waveguides with quantum dot
  molecules. \emph{Phys. Rev. B} \textbf{2019}, \emph{99}, 085311\relax
\mciteBstWouldAddEndPuncttrue
\mciteSetBstMidEndSepPunct{\mcitedefaultmidpunct}
{\mcitedefaultendpunct}{\mcitedefaultseppunct}\relax
\EndOfBibitem
\bibitem[Cygorek \latin{et~al.}(2020)Cygorek, Korkusinski, and
  Hawrylak]{ref:Marek1}
Cygorek,~M.; Korkusinski,~M.; Hawrylak,~P. Atomistic theory of electronic and
  optical properties of {I}n{A}s{P}/{I}n{P} nanowire quantum dots. \emph{Phys.
  Rev. B} \textbf{2020}, \emph{101}, 075307\relax
\mciteBstWouldAddEndPuncttrue
\mciteSetBstMidEndSepPunct{\mcitedefaultmidpunct}
{\mcitedefaultendpunct}{\mcitedefaultseppunct}\relax
\EndOfBibitem
\bibitem[Vdovin \latin{et~al.}(2000)Vdovin, Levin, Patan\`{e}, Eaves, Main,
  Khanin, Dubrovskii, and Hill]{ref:Map1}
Vdovin,~E.~E.; Levin,~A.; Patan\`{e},~A.; Eaves,~L.; Main,~P.~C.;
  Khanin,~Y.~N.; Dubrovskii,~Y.~V.; Hill,~M. H.~G. Imaging the Electron Wave
  Function in Self-Assembled Quantum Dots. \emph{Science} \textbf{2000},
  \emph{290}, 122\relax
\mciteBstWouldAddEndPuncttrue
\mciteSetBstMidEndSepPunct{\mcitedefaultmidpunct}
{\mcitedefaultendpunct}{\mcitedefaultseppunct}\relax
\EndOfBibitem
\bibitem[Patan\`{e} \latin{et~al.}(2002)Patan\`{e}, Hill, Eaves, Main, Henini,
  Zambrano, Levin, Mori, Hamaguchi, Dubrovskii, Vdovin, Austing, Tarucha, and
  Hill]{ref:Map2}
Patan\`{e},~A.; Hill,~R. J.~A.; Eaves,~L.; Main,~P.~C.; Henini,~M.;
  Zambrano,~M.~L.; Levin,~A.; Mori,~N.; Hamaguchi,~C.; Dubrovskii,~Y.~V.;
  Vdovin,~E.~E.; Austing,~D.~G.; Tarucha,~S.; Hill,~G. Probing the quantum
  states of self-assembled {I}n{A}s dots by magnetotunneling spectroscopy.
  \emph{Phys. Rev. B} \textbf{2002}, \emph{65}, 165308\relax
\mciteBstWouldAddEndPuncttrue
\mciteSetBstMidEndSepPunct{\mcitedefaultmidpunct}
{\mcitedefaultendpunct}{\mcitedefaultseppunct}\relax
\EndOfBibitem
\bibitem[Reimer \latin{et~al.}(2011)Reimer, {van Kouwen}, Hidma, {van Weert},
  Bakkers, Kouwenhoven, and Zwiller]{ReimerNL}
Reimer,~M.~E.; {van Kouwen},~M.~P.; Hidma,~A.~W.; {van Weert},~M. H.~M.;
  Bakkers,~E. P. A.~M.; Kouwenhoven,~L.~P.; Zwiller,~V. Electric Field Induced
  Removal of the Biexciton Binding Energy in a Single Quantum Dot. \emph{Nano
  Letters} \textbf{2011}, \emph{11}, 645\relax
\mciteBstWouldAddEndPuncttrue
\mciteSetBstMidEndSepPunct{\mcitedefaultmidpunct}
{\mcitedefaultendpunct}{\mcitedefaultseppunct}\relax
\EndOfBibitem
\bibitem[Fasth \latin{et~al.}(2005)Fasth, Fuhrer, Bj{\"o}rk, and
  Samuelson]{Fasth}
Fasth,~C.; Fuhrer,~A.; Bj{\"o}rk,~M.~T.; Samuelson,~L. Tunable Double Quantum
  Dots in InAs Nanowires Defined by Local Gate Electrodes. \emph{Nano Letters}
  \textbf{2005}, \emph{5}, 1487\relax
\mciteBstWouldAddEndPuncttrue
\mciteSetBstMidEndSepPunct{\mcitedefaultmidpunct}
{\mcitedefaultendpunct}{\mcitedefaultseppunct}\relax
\EndOfBibitem
\end{mcitethebibliography}

\end{document}